\documentclass[
letterpaper,
reprint,           
twocolumn,
superscriptaddress,
amsmath,           
amssymb,           
aps,               
prl,               
showkeys,          
showpacs,          
notitlepage,       
longbibliography,  
floatfix,          
nofootinbib,
]{revtex4-1}
\usepackage{bm}
\usepackage{amsmath}
\usepackage{lipsum}
\allowdisplaybreaks[4]
\usepackage{cancel}
\usepackage{extarrows}
\usepackage{tensor}     
\usepackage{float}
\usepackage[caption = false]{subfig} 
\usepackage[final]{graphicx}   
\usepackage[
colorlinks=true,        
citecolor=blue,         
linkcolor=blue,         
urlcolor=blue           
]{hyperref}             
\usepackage{bm}         
\usepackage{xcolor}     
\usepackage{lipsum}
\usepackage{color}      
\usepackage[utf8]{inputenc} 
\usepackage[section]{placeins} 
\usepackage{appendix}
\usepackage{units}
\usepackage[capitalise]{cleveref}
\usepackage{units}
\newcommand{\nc}{\newcommand*}

\nc{\xbar}{\bar{x}}
\nc{\rhoeq}{\rho_{\mathrm{eq}}}
\nc{\zeq}{z_{\mathrm{eq}}}
\nc{\tla}{\tilde{\lambda}}
\nc{\bt}{\beta}
\nc{\dt}{\delta}
\nc{\Dt}{\Delta}
\nc{\vj}{\vec{j}}
\nc{\vl}{\vec{l}}
\nc{\hx}{\hat{x}}
\nc{\hy}{\hat{y}}
\nc{\bj}{\bm{j}}
\nc{\mJ}{\mathcal{J}}
\nc{\mP}{\mathcal{P}}
\nc{\Msun}{M_\odot}
\nc{\Mc}{\mathcal{M}}
\nc{\app}{\approx}
\nc{\av}[1]{\langle #1 \rangle}
\nc{\eq}[1]{Eq.~\eqref{#1}}
\nc{\al}{\alpha}
\nc{\Xstar}{X_{\ast}}
\nc{\fpbh}{f_{\mathrm{pbh}}}
\nc{\vth}{\vec{\theta}}
\nc{\vla}{\vec{\lambda}}
\nc{\vd}{\vec{d}}
\nc{\Mmin}{M_{\mathrm{min}}}
\nc{\rmd}{\mathrm{d}}
\nc{\mmin}{{m_{\mathrm{min}}}}
\nc{\mmax}{{m_{\mathrm{max}}}}
\nc{\mR}{\mathcal{R}}
\nc{\tmR}{\tilde{\mathcal{R}}}
\nc{\s}{\sigma}
\nc{\ogw}{\Omega_{\mathrm{GW}}}
\nc{\addref}{[\textcolor{red}{add ref}] }
\nc{\Om}{\Omega}
\nc{\gm}{\gamma}
\nc{\Gm}{\Gamma}
\nc{\gpcyr}{\mathrm{Gpc}^{-3}\,\mathrm{yr}^{-1}}
\nc{\Eq}[1]{Eq.~\eqref{#1}}
\nc{\Fig}[1]{Fig.~\ref{#1}}
\nc{\Table}[1]{Table~\ref{#1}}
\nc{\lvc}{LIGO/Virgo} 
\nc{\Sec}[1]{Sec.~\ref{#1}}
\nc{\eg}{\textit{e.g.~}}
\nc{\SNR}{\mathrm{SNR}}
\nc{\be}{\mathbf{\epsilon}}
\nc{\bn}{\mathbf{n}}
\nc{\bd}{\mathbf{d}}
\nc{\ba}{\mathbf{a}}
\nc{\eps}{\epsilon}
\nc{\bnu}{\mathbf{\nu}}
\nc{\mb}{\mathbf}
\nc{\bbt}{\mathbf{t}}
\nc{\bth}{\mathbf{\theta}}
\nc{\bep}{\mathbf{\epsilon}}
\nc{\uni}{\mathrm{U}}
\nc{\logu}{\operatorname{\mathrm{log-U}}}
\nc{\RN}{\mathrm{RN}}
\nc{\BN}{\mathrm{BN}}
\nc{\GN}{\mathrm{GN}}
\nc{\mcN}{\mathcal{N}}
\nc{\GWB}{\mathrm{GW}}
\nc{\yr}{\mathrm{yr}}
\nc{\Am}{\mathcal{A}}
\nc{\Dm}{\mathcal{D}}
\nc{\Hm}{\mathcal{H}}

\nc{\mrm}{\mathrm}
\nc{\BE}{B\scriptsize{AYES}\normalsize{E}\scriptsize{PHEM}\normalsize  }

\nc{\Ostgw}{\Omega_{\mathrm{GW}}^{\mathrm{ST}}}
\nc{\Ottgw}{\Omega_{\mathrm{GW}}^{\mathrm{TT}}}
\nc{\Ovlgw}{\Omega_{\mathrm{GW}}^{\mathrm{VL}}}
\nc{\Oslgw}{\Omega_{\mathrm{GW}}^{\mathrm{SL}}}
\nc{\cosxi}{\beta}

\nc{\gmPL}{\gamma_{\mathrm{PL}}}
\nc{\APL}{A_{\mathrm{PL}}}

\def\({\left(}
\def\){\right)}
\def\[{\left[}
\def\]{\right]}

\def\e{\begin{equation}}
\def\q{\end{equation}}
\def\m{\begin{eqnarray}}
\def\n{\end{eqnarray}}
\nc{\red}[1]{\textcolor{red}{#1}}

\begin{document}

\title{Implications for the Supermassive Black Hole Binaries from the NANOGrav 15-year Data Set}

\author{Yan-Chen Bi}
\email{biyanchen@itp.ac.cn} 
\affiliation{CAS Key Laboratory of Theoretical Physics, 
    Institute of Theoretical Physics, Chinese Academy of Sciences,Beijing 100190, China}
\affiliation{School of Physical Sciences, University of Chinese Academy of Sciences, No. 19A Yuquan Road, Beijing 100049, China}
    
\author{Yu-Mei Wu}
\email{Corresponding author: ymwu@ucas.ac.cn} 
\affiliation{School of Fundamental Physics and Mathematical Sciences, Hangzhou Institute for Advanced Study, UCAS, Hangzhou 310024, China}
\affiliation{School of Physical Sciences, 
    University of Chinese Academy of Sciences, 
    No. 19A Yuquan Road, Beijing 100049, China}

\author{Zu-Cheng~Chen}
\email{Corresponding author: zucheng.chen@bnu.edu.cn}
\affiliation{Department of Astronomy, Beijing Normal University, Beijing 100875, China}
\affiliation{Advanced Institute of Natural Sciences, Beijing Normal University, Zhuhai 519087, China}
\affiliation{Department of Physics and Synergistic Innovation Center for Quantum Effects and Applications, Hunan Normal University, Changsha, Hunan 410081, China}

\author{Qing-Guo Huang}
\email{Corresponding author: huangqg@itp.ac.cn}
\affiliation{School of Fundamental Physics and Mathematical Sciences, Hangzhou Institute for Advanced Study, UCAS, Hangzhou 310024, China}
\affiliation{CAS Key Laboratory of Theoretical Physics, 
    Institute of Theoretical Physics, Chinese Academy of Sciences,Beijing 100190, China}
\affiliation{School of Physical Sciences, 
    University of Chinese Academy of Sciences, 
    No. 19A Yuquan Road, Beijing 100049, China}


\begin{abstract}
Several pulsar timing array (PTA) collaborations, including NANOGrav, EPTA, PPTA, and CPTA,  have announced the evidence for a stochastic signal consistent with a stochastic gravitational wave background (SGWB). Supermassive black hole binaries (SMBHBs) are supposed to be the most promising  gravitational-wave (GW) sources for this signal. In this letter, we use the NANOGrav 15-year data set to constrain the parameter space in  an astro-informed formation model for SMBHBs. Our results prefer a large turn-over eccentricity of the SMBHB orbit when GWs begin to dominate the SMBHB evolution.
Furthermore, the SGWB spectrum is extrapolated to the space-borne GW detector frequency band by including inspiral-merge-cutoff phases of SMBHBs, indicating that the SGWB from SMBHBs should be detected by LISA, Taiji and TianQin in the near future. 
\end{abstract}
\keywords{supermassive black hole binary, gravitational wave, pulsar timing array}
\pacs{04.30.Db, 04.80.Nn, 95.55.Ym}
\maketitle

\textbf{Introduction.} 
Pulsar timing arrays (PTAs)~\citep{1990ApJ:PTA}, comprised of a set of millisecond pulsars, are proposed to detect the gravitational waves (GWs) at nHz. 
Recently, the North American Nanohertz Observatory for Gravitational Waves (NANOGrav)~\citep{NANOGrav:2023,NANOGrav:2023evidence}, the European Pulsar Timing Array (EPTA) align with the Indian Pulsar Timing Array (InPTA)~\citep{Antoniadis:2023EPTA1,Antoniadis:2023EPTA3}, the Parkes Pulsar Timing Array (PPTA)~\citep{Zic:2023PPTA1, Reardon:2023PPTA2} and the Chinese Pulsar Timing Array (CPTA)~\citep{Xu:2023CPTA} have announced evidence for a stochastic signal consistent with the Hellings-Downs~\citep{Hellings:1983HD} correlations, pointing to the stochastic gravitational-wave background (SGWB) origin of this signal. Although some exotic new physics can generate GWs in PTA frequency band~\cite{Kibble:1976sj, Vilenkin:1984ib,Damour:2004kw,Siemens:2006yp,Caprini:2010xv,Kobakhidze:2017mru,Arunasalam:2017ajm,Xue:2021gyq,Maggiore:1999vm,Saito:2008jc,Caprini:2018mtu,Chen:2019xse,Cai:2019bmk,Yuan:2019fwv,Li:2019vlb,Yuan:2019wwo,Yuan:2019udt,Liu:2021jnw,Chen:2021ncc,Moore:2021ibq,Wu:2021kmd,Chen:2021wdo,PPTA:2022eul,Chen:2022azo,Bian:2022tju,Meng:2022low,Chen:2021nxo,NANOGrav:2023hvm,Wu:2023dnp,Wu:2023pbt,Wu:2023hsa,Liu:2023ymk,Guo:2023hyp,Broadhurst:2023tus,Yang:2023aak,Konoplya:2023fmh,Lazarides:2023ksx,King:2023cgv,Niu:2023bsr,Jin:2023wri,Madge:2023cak,Zhu:2023faa,Wang:2023ost,Li:2023qua,Wang:2023div,Addazi:2023jvg,Franciolini:2023pbf,Ellis:2023tsl,Bian:2023dnv,Liu:2023pau,Yi:2023npi,Yi:2023tdk,You:2023rmn,IPTA:2023ero,Chen:2023zkb,InternationalPulsarTimingArray:2023mzf}, supermassive black hole binaries (SMBHBs) are widely regarded as the most promising GW source for this signal~\citep{NANOGrav:2023SMBHBGWB, NANOGrav:2023SMBHGW,Antoniadis:2023EPTA5,Ellis:2023dgf,Li:2023awp}.


SMBHBs are supposed to form after the mergers of galaxies, but the detailed process of their formation, evolution, and correlation with host galaxies remains unresolved. In the scenario of galaxy coalescence, supermassive black holes (SMBHs) hosted in their nuclei sink to the center of the remnant due to the interaction with the surrounding ambient and eventually form bound SMBHBs \citep{Begelman:1980smbhevo}. The SMBHBs subsequently harden and may tend to increase the binary eccentricity because of dynamical interaction with the dense background \citep{Dotti:2011um} and scattering of ambient stars until gravitational waves (GWs) take over at sub-parsec separation \citep{Kocsis:2010inter, Sesana:2013gxtobh, Ravi:2014inter, Kelley:2016inter, Chen:2016spectrum, Rodig:2011ecc, Quinlan:1996ecc, Armitage:2005ecc, Chen:2018galaxy18}. Since galaxies are observed to merge quite frequently \citep{Bell:2006merger, deRavel:2008merger} and the observable Universe encompasses several billions of them, a largish cosmological population of SMBHBs is expected to produce the SGWB \citep{Sesana:2008merger}. 
This complicated process makes it difficult to be modeled accurately. However, 
it is widely recognized that galaxies have a correlation with SMBHs at their centers \citep{KormendyHo:2013}. This correlation can be used as an astrophysical model to describe the merger rate of SMBHBs, known as the astro-informed formation model \citep{Chen:2016kax,Chen:2016spectrum,Chen:2018galaxy18}. 

During the evolution of SMBHBs, the inspiral phase falls within the PTA frequency range, while the merger phase falls within the sensitive range of space-borne GW observatories, such as LISA \citep{LISA:2017}, Taiji \citep{Taiji:2021}, and TianQin \citep{TianQin:2015}, operating in the frequency band of approximately $10^{-4} \sim 10^{-1}$ Hz. By observing SGWB signal from PTA, we can unravel the evolution of SMBHBs  and then shed light on the signals for the space-borne GW detectors.

\begin{figure}[htbp]
	\centering
	\includegraphics[width=0.5\textwidth]{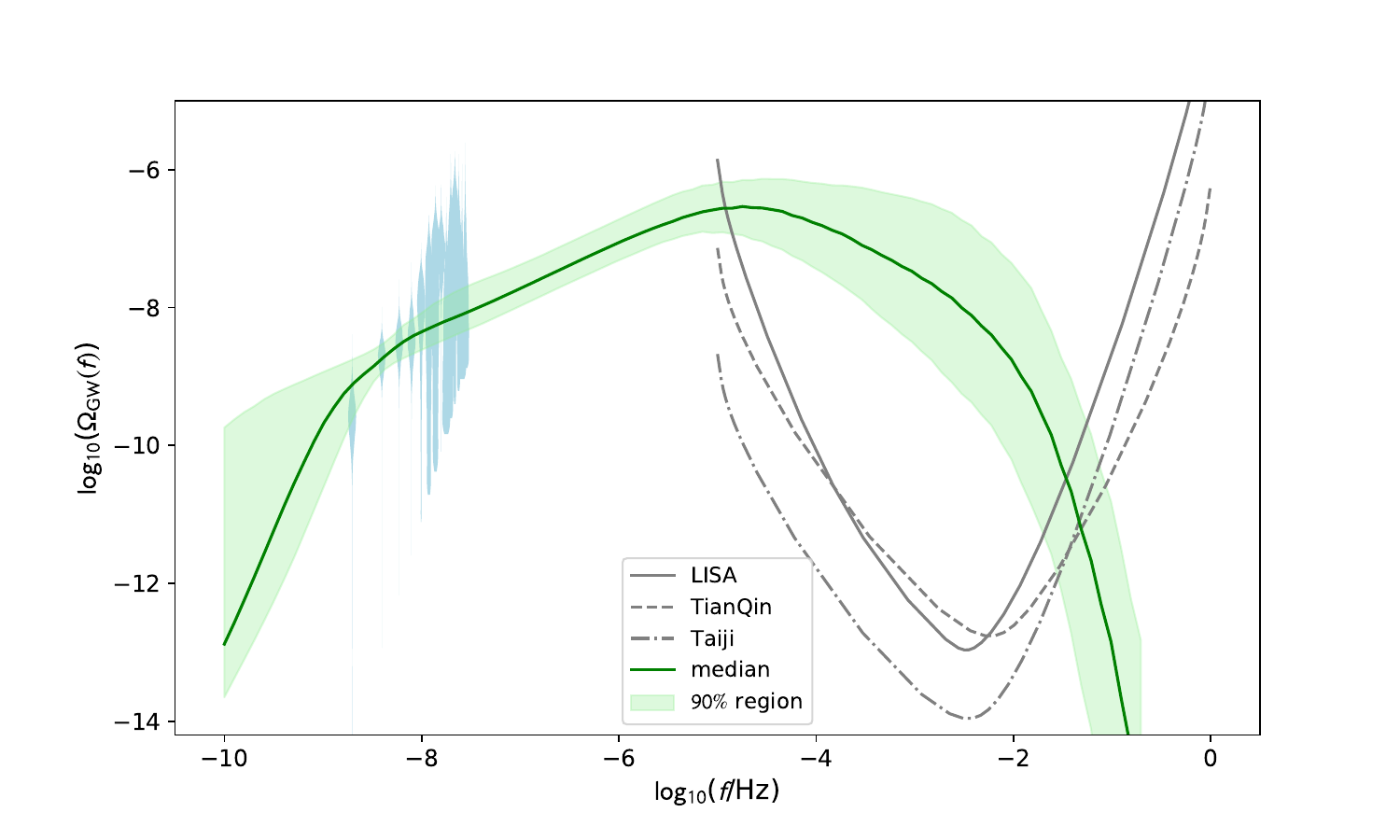}
	\caption{\label{spect} The expected SGWB spectrum $\Omega_{\rm{GW}}(f)$ from the SMBHBs in the frequency range $[10^{-10},10^{-1}] \rm{Hz}$. The light-blue violin plots show the posterior of the first 14-frequency bins for the NANOGrav 15-year data set. The green line shows the medium spectrum obtained from the NANOGrav analysis based on the astro-informed formation model, and the light-green shaded region indicates the $90\%$ credible interval. We also show the expected power-law integrated curves of LISA, TianQin and Taiji with solid, dashed, dash-dotted gray lines, respectively.}
\end{figure}

In this letter, we utilize the NANOGrav 15-year data set to constrain the astro-informed formation model of SMBHBs by including the relatively strong constraints naturally provided by astronomical surveys \citep{Conselice:2016GSMF, Sesana:2013gxtobh, Kormendy:2013gxtobh, Chen:2018galaxy18}. We find that the orbits of SMBHBs exhibit significant eccentricity when GWs begin to dominate their evolution. Furthermore, we extrapolate the SGWB spectrum from $10^{-9}$ Hz to $10^{-1}$ Hz (see \Fig{spect}), highlighting the potential of SMBHBs as a promising source for space-borne GW detectors.




\textbf{SGWB from SMBHB.} 
The SGWB from SMBHBs is the most promising target of PTA. In general SMBHB is supposed to form following the merge of two galaxies. After galaxies merging, their central SMBHs sink into the center of the merger remnant and form a bound binary. Initially the binary orbit shrinks due to the energy and angular momentum exchanging with surrounding stars and cold gas. Then GW radiations will dominate the evolution at turn-over frequency $f_t$ with initial eccentricity $e_0$, bringing the binary to final coalescence~\citep{Dotti:2011SMBHB}. The spectrum of SGWB is composed from the sum of all GWs emitted by SMBHBs at a given observed frequency $f$. The present-day energy density of SGWB $\ogw(f)$ is given by~\citep{Phinney2001GWB}
\begin{equation}
    \ogw(f) = \frac{8 \pi G f }{3 H_0^2 c^2} \int dz d\Mc \frac{d^2 n}{dz d\Mc} \frac{dE_{\rm GW}}{df_r},
    \label{ogw2}
\end{equation}
where $H_0 = 67.4 \ {\rm km\ sec^{-1}\ Mpc^{-1}}$ \citep{Planck:2018vyg} is the Hubble constant, $f_r = (1+z)f$ is the source-frame GW frequency. Here, $\Mc = M q^{3/5} / (1 + q)^{1/5}$ is the chirp mass of SMBHB, where $M$ is the primary mass and $q$ is the mass ratio. Also, $d^2n/dz d\Mc$ is the SMBHB population
and ${dE_{\rm GW}}/{df_r}$ is the energy spectrum of a  single SMBHB. The relevant ranges in the integrals used here are $0 \le z \le 5$ and $10^5 \Msun \le \Mc \le 10^{11} \Msun$. 

The SMBHB population ${d^2 n}/{dz d\Mc}$ in \Eq{ogw2} are estimated from astrophysical observation~\citep{Sesana:2008galaxymodel}
\m
\frac{d^2 n}{dz^{\prime} d\Mc} = \int \frac{d^3 n_G}{dz^{\prime} dM_G dq_G} \frac{dM_G}{dM} \frac{dq_G}{dq} \frac{dM}{d\Mc} dq,
\n
where $M_G$ is the primary galaxy mass, $q_G$ is the mass ratio between the two paired galaxy, and $q = q_G^{\alpha_*}$ is integrated from $0.25$ to $1$. The galaxy mass $M_{G}$ and the black hole mass $M$ are related through the bulge mass $M_{\rm{bulge}}$ \citep{Sesana:2013gxtobh, Kormendy:2013gxtobh},
\begin{widetext}
\begin{equation}
\frac{M_{\rm bulge}}{M_G} = \left\{ 
\begin{array}{lr}
    \frac{\sqrt{6.9}}{(\log_{10} M_G - 10)^{1.5}} \exp\(\frac{-3.45}{\log_{10} M_G - 10}\) + 0.615 &\quad {\rm if} \log_{10} M_G \ge 10, \\
    0.615 &\quad {\rm if} \log_{10} M_G \le 10.
\end{array} \right.   
\label{MbulgeMg}
\end{equation}
\end{widetext}
Note that this relation is only valid for red galaxy pairs which completely dominated the SGWB in PTA bands \citep{Sesana:2013gxtobh}. The primary mass $M$ is given by
\e
M = \mathcal{N}\left\{ M_*\(\frac{M_{\rm bulge}}{10^{11} \Msun} \), \epsilon \right\},
\label{MMbulge}
\q
with $\mathcal{N}\{x,y\}$ a log normal distribution with mean $x$ and standard deviation $y$. Here $\epsilon$ is the scatter for $M_{\rm bulge}$-$M_{\rm BH}$ relation.

The galaxy differential merger rate per unit redshift, galaxy mass and mass ratio ${d^3n_G}/{dz^{\prime} dM_G dq_G}$ is~\citep{Sesana:2008galaxymodel}
\begin{equation}
     \frac{d^3n_G}{dz^{\prime} dM_G dq_G} = \frac{\Phi(M_G,z)}{M_G \ln 10} \frac{\mathcal{F}(M_G,z,q_G)}{\tau(M_G,z,q_G)} \frac{dt_r}{dz},
     \label{galaxy_model}
\end{equation}
where $z$ stands for the redshift of galaxy pair, $z^{\prime}$ is the redshift at galaxy pair merging, and they are connected by merger timescale of the pair galaxy $\tau(M_G,z,q_G)$. The quantitative relation between $z$ and $z^{\prime}$ is addressed as \citep{Chen:2018galaxy18}
\e
\int_{z^{\prime}}^{z} \frac{dt}{d\tilde{z}} d\tilde{z} = \tau(M_G,z,q_G).
\q
Furthermore, $dt_r/dz$ is the relationship between time and redshift assuming a flat $\Lambda$CDM Universe
\e
\frac{dt_r}{dz} = \frac{1}{H_0(1+z)\sqrt{\Omega_M(1+z)^3  + \Omega_{\Lambda}}},
\q
and $\Phi(M_G,z) \equiv {dn_G}/{d \log_{10} M_G}$ is a single Schechter function describing the galaxy stellar mass function (GSMF) measured at redshift $z$ for galaxy pair, and is given by \citep{Mortlock:2014GSMF, Conselice:2016GSMF} 
\e
\Phi(M_G,z) = 10^{\Phi(z)} (\ln 10) \(\frac{M_G}{M_{G0}}\)^{1 + \alpha(z)} \exp{\(-\frac{M_G}{M_{G0}}\)},
\label{GSMF}
\q
where $\Phi(z) = \Phi_0 + z \Phi_I$, and $\alpha(z) = \alpha_0 + z \alpha_I$.
The differential pair fraction with respect to the mass ratio of galaxy pair $q_G$ is \citep{Chen:2018galaxy18}
\e
\mathcal{F}(M_G,z,q_G) = \frac{df_{\rm pair}}{dq_G} = f_0^{\prime} \( \frac{M_G}{a M_{G0} } \)^{\alpha_f} \(1+z\)^{\beta_{f}} q_G^{\gamma_f},
\label{pair}
\q
where $aM_{G0} = 10^{11} \Msun$ is an arbitrary reference mass~\citep{Chen:2018galaxy18}. Here, $f_0^{\prime}$ is related to $f_0$ via $f_0 = f_0^{\prime} \int q_G^{\gamma_f} dq_G$. 
The merger timescale of the pair galaxy in Eq.~\eqref{galaxy_model}, $\tau(M_G,z,q_G)$, can be expressed as \citep{Snyder:2017timescale}
\e
\tau(M_G,z,q_G) = \tau_0 \(\frac{M_G}{b M_{G0}} \)^{\alpha_{\tau}} \( 1+z \)^{\beta_{\tau}} q_G^{\gamma_{\tau}},
\label{timescale}
\q
where $\tau_0$ represents merger time normalization with unit of $\rm{Gyr}$. Meanwhile, $bM_{G0} = 0.4/h_0 \times 10^{11} \Msun$ is an arbitrary reference mass. Noted the merger timescale describes the time elapsed between the observed galaxy pair and the final coalescence of the SMBHB, including the time for two galaxies to effectively merge, and the time required for the SMBHs to form a binary and merge \citep{Chen:2018galaxy18}.

The energy spectrum of a single SMBHB $dE_{\rm GW}/df_r$ in \Eq{ogw2} can be calculated using its self-similarity. In other words, a purely GW emission-driven binary
spectrum in any configuration can be obtained from a reference spectrum by shift and re-scaling~\citep{Chen:2016spectrum}.
The fiducial redshift and chirp mass of reference spectrum are set as $z_0 = 0.02$ and $\Mc_0 = 4.16 \times 10^{8} \Msun$, respectively. 


Here, we consider a realistic insprial phase that evolves with an initial eccentricity, and the inspiral-merger-cutoff phases are jointed smoothly following methods proposed by \citep{ Poisson:1993dedf, Finn:1992dedf, Zhu:2011dedf, Chen:2016spectrum}. We express the complete description of energy spectrum $dE_{\rm GW}/{df_r}$ as follow 
\m
\frac{dE_{\rm GW}}{df_r}(f_r<\nu_1) &=& \frac{\pi c^2 f}{4 G} h^2_{\rm c, fit}\(f \frac{f_{\rm p,\rm{ref}}}{f_{\rm p,t}} \) \(\frac{f_{\rm p,\rm{ref}}}{f_{\rm p,t}} \)^{-\frac{4}{3}} \nonumber \\ &&\times \(\frac{\Mc}{\Mc_0} \)^{\frac{5}{3}} \(\frac{1 + z}{1 + z_0} \)^{-\frac{1}{3}},  \label{dedf1} \\
\frac{dE_{\rm GW}}{df_r}(f_r\in [\nu_1,\nu_2)) &=& \frac{(G \pi)^{2/3}\Mc^{5/3}}{3}\omega_1 f_r^{2/3}, \label{dedf2} \\
\frac{dE_{\rm GW}}{df_r}(f_r\in [\nu_2,\nu_3)) &=& \frac{(G \pi)^{2/3}\Mc^{5/3}}{3} \omega_2\left[\frac{f_r}{1+(\frac{f_r-\nu_{2}}{\sigma/2})^2}\right]^2, \nonumber \label{dedf3} \\
\n
where $\omega_1=\nu_1^{-1}$, and $\omega_2=\nu_1^{-1}\nu_2^{-4/3}$. The set of parameters $(\nu_1,\nu_2,\sigma,\nu_3)$ can be determined by the total mass $M_{\rm total} = M (1+q)$ and the symmetric mass ratio $\eta=(q M^2/M_{\rm total}^2)$ in terms of  $(a \eta^2+b\eta+c)/(\pi GM_{\rm total}/c^3)$, with coefficients $a,b,c$ given in Table 1 of ~\citep{Ajith:2007kx}. The ratio ${f_{\rm p,ref}}/{f_{\rm p,t}}$ for shift is \citep{Chen:2016spectrum}
\e
    \frac{f_{\rm p,{\rm ref}}}{f_{\rm p,t}} = \frac{f_{\rm ref}}{f_t} \[ \(\frac{e_{\rm ref}}{e_0}\)^{\frac{12}{19}} \frac{1 - e_0^2}{1 - e_{\rm ref}^2}  \left( \frac{ 304 + {121 e_{\rm ref}^2} } { 304 + {121 e_0^2}  }\right)^{\frac{870}{2299}} \]^{\frac{3}{2}},
\q
where $f_{\rm ref} = 10^{-10} {\rm Hz}$ and $e_{\rm ref} = 0.9$ are the reference frequency and ecctricity, respectively. 
Here, $e_0$ is the initial eccentricity when GWs begin to dominate and $f_t$ is the turn-over frequency given by
\e
f_t = 0.356 {\rm nHz} \(\frac{\rho_{ i,100}}{F(e_0) \sigma_{200}} \zeta_0 \)^{\frac{3}{10}} \Mc_9^{-\frac{2}{5}},
\label{ft}
\q
where 
\e
F(e) = \frac{1 + (73/24)e^2 + (37/96)e^4}{(1-e^2)^{7/2}},
\q
$\Mc_9 = \Mc/(10^9 \Msun)$ is the rescaled chirp mass, $\rho_{i,100} = \rho_i / (100\, \Msun \, \mathrm{pc}^{-3}) $ is  density of the stellar environment at the influence radius of the SMBHB, $\sigma_{200}=\sigma/(200\, \rm{km/s})$ is the velocity dispersion of stars in the galaxy, and $\zeta_0$ is an additional multiplicative factor absorbing all systematic uncertainties in the estimate of $\rho_{i,100}$ and $\sigma_{200}$. Note that massive galaxies generally have velocity dispersions in the range of $250\, \mathrm{km/s} < \sigma < 350\, \mathrm{km/s}$, and such a narrow range has a negligible influence on the results.
The analytical fitting function of spectrum, $h^2_{\rm c, fit}(f)$, takes the form as \citep{Chen:2016spectrum}
\begin{equation}
    h_{\rm c, fit}(f) = a_0 \bar{f}^{a_1} e^{-a_2 \bar{f}} + b_0 \bar{f}^{b_1} e^{-b_2 \bar{f}} + c_0 \bar{f}^{- c_1} e^{-c_2 / \bar{f}},
    \label{hcfit}
\end{equation}
where $\bar{f} = f/(10^{-8}) {\rm Hz}$, and the values of $a_i, b_i, c_i$ can given below Equation 15 in Ref.~\citep{Chen:2016spectrum}.


To sum up, the present-day energy density $\ogw(f)$ of SGWB from SMBHBs in astro-informed formation model is fully specified by a set of eighteen model parameters: 
$\{\Phi_0, \Phi_I, M_{G0}, \alpha_0, \alpha_I\}$ for the GSMF, $\{f_0, \alpha_f, \beta_{f}, \gamma_f\}$ for the pair fraction, $\{\tau_0, \alpha_{\tau}, \beta_{\tau}, \gamma_{\tau}\} $ for the merger timescale, $\{ M_*, \alpha_*,\epsilon \} $ for galaxy-SMBH transforming relation and $\{e_0, \zeta_0 \}$ for single SMBHB energy spectrum. The detailed descriptions of parameters are addressed in Table~\ref{tab:galaxypara}.

\begin{table}[htbp]
    \setlength{\tabcolsep}{3pt}
    \caption{List of the 18 parameters and their prior in the astro-informed formation model. Here U represents a uniform distribution.}
    \begin{tabular}{clc}
    \hline\hline
    parameter  &  description  &  prior \\
    \hline
             & \textbf{GSMF}  & \\
    $\Phi_0$ & GSMF norm & $-2.77_{-0.29}^{+0.27}$  \\
    $\Phi_I$ & GSMF norm redshift evolution & $-0.27_{-0.21}^{+0.23}$   \\
    $\log_{10} M_{G0}$ & GSMF scaling mass & $11.24_{-0.17}^{+0.20}$  \\
    $\alpha_0$ & GSMF mass slope   & $-1.24_{-0.16}^{+0.16}$   \\
    $\alpha_I$ & GSMF mass slope redshift evolution   & $-0.03_{-0.14}^{+0.16}$    \\
    \hline 
          & \textbf{Galaxy pair function} & \\
    $f_0$ & pair fraction norm  & U$[0.01,0.05]$    \\
    $\alpha_f$ & pair fraction mass slope  & U$[-0.5,0.5]$     \\
    $\beta_f$ & pair fraction redshift slope  & U$[0,2]$   \\
    $\gamma_f$ & pair fraction mass ratio slope  & U$[-0.2,0.2]$   \\
    \hline 
           & \textbf{Galaxy merger timescale} & \\
    $\tau_0$ & merger time norm & U$[0.1, 10]$\\
    $\alpha_{\tau}$ & merger time mass slope   & U$[-0.5,0.5]$     \\
    $\beta_{\tau}$ & merger time redshift slope  & U$[-3,1]$     \\
    $\gamma_{\tau}$ & merger time mass ratio slope  & U$[-0.2,0.2]$\\
    \hline 
         & \textbf{$\bm{M_{\rm bulge}}$-$\bm{M_{\rm BH}}$ relation}     &  \\
    $\log_{10} M_{*}$ & $M_{\rm bulge}$-$M_{\rm BH}$ relation norm & $8.17_{-0.32}^{+0.35}$\\
    $\alpha_*$ & $M_{\rm bulge}$-$M_{\rm BH}$ relation slope  & $1.01_{-0.10}^{+0.08}$\\
    $\epsilon$ & $M_{\rm bulge}$-$M_{\rm BH}$ relation scatter  & U$[0.2,0.5]$\\
    \hline 
         & \textbf{Stellar and SMBHB condition}  & \\
    $e_0$ &  SMBHB initial eccentricity   & U$[0.01,0.99]$\\
    $\log_{10} \zeta_0$ & stellar density factor   & U$[-2,2]$\\
    \hline
    \end{tabular} 
    \label{tab:galaxypara}
\end{table}

\begin{figure*}[htbp!]
	\centering
	\includegraphics[width=\textwidth]{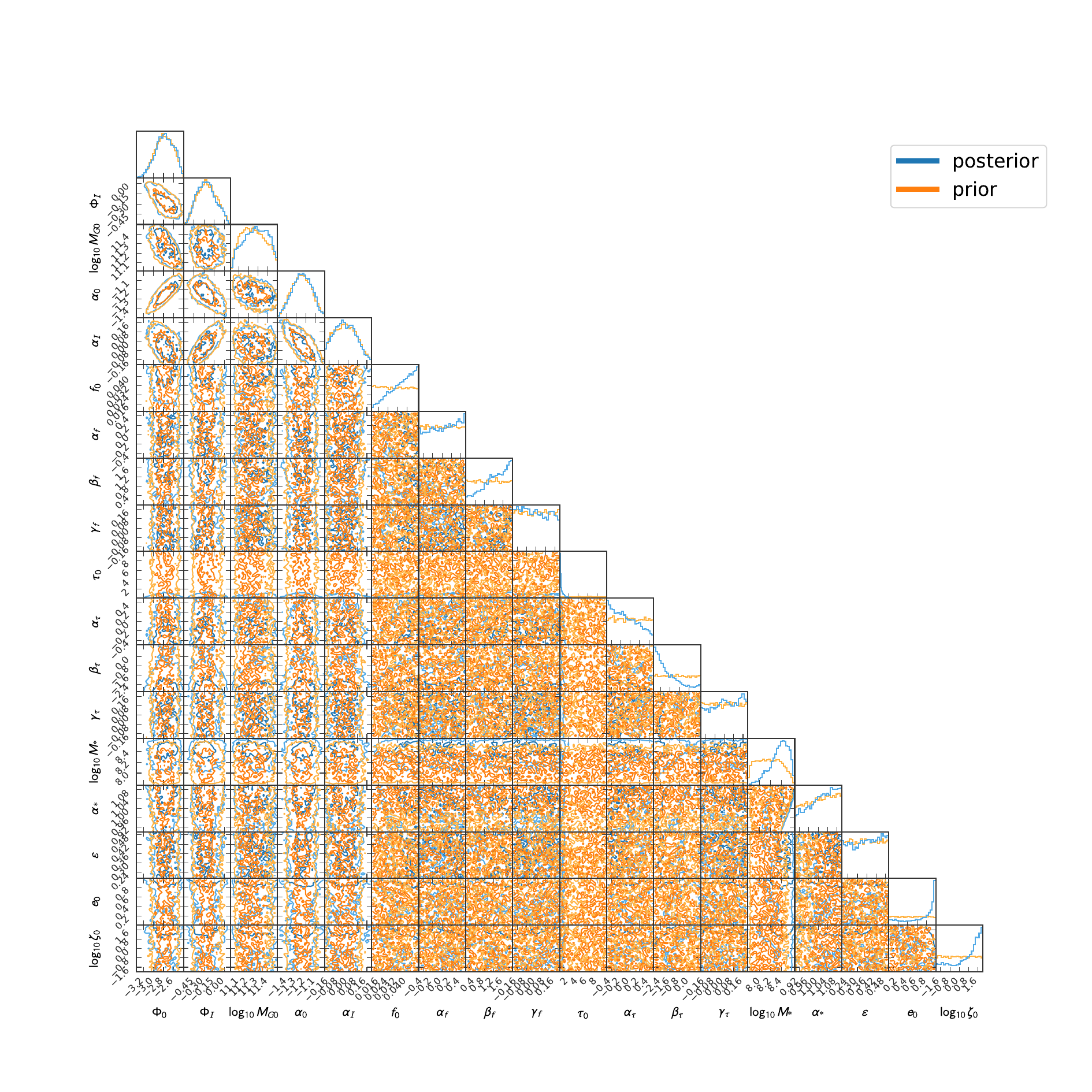}\caption{\label{post_18params} One- and two-dimensional prior distribution (in the orange color) and  marginalised posterior distribution (in the blue color) for the 18 parameters in the astro-informed formation model. We show the $1\sigma$ and $2\sigma$ contours in the two-dimensional plots.  }
\end{figure*}

\textbf{Data and results.} 
The NANOGrav collaboration has performed an analysis on their 15-year data set by employing a free spectrum that enables independent variations in the amplitude of the GW spectrum across different frequencies. In the analyses, we use the posterior data from NANOGrav \citep{NANOGrav:2023evidence,NANOGrav:2023newphysics}, and the \texttt{PTMCMCSampler} \citep{justin_ellis_2017_1037579} package to perform the Markov Chain Monte Carlo sampling. We note that the constrained prior distributions for certain model parameters in \Table{tab:galaxypara} are based on those presented in \citep{Chen:2018galaxy18} and are derived from observational and theoretical works on the measurement of the GSMF, galaxy pair fraction, merger timescale and SMBH-host galaxy scaling relations.

The resulting posterior distribution is illustrated in \Fig{post_18params}. Some parameters are prior dominated. However, our analysis also reveals that the detected stochastic signal provides some new insights into the differential pair fraction $\mathcal{F}(M_{G},z,q_{G})$, merger timescale $\tau({M_{G}},z,q_{G})$, galaxy-SMBH mass scaling $M_G-M_{\rm{bulge}}-M$ relation, and the initial states of the SMBHBs when GW emission takes over, such as the eccentricity $e_{0}$ and the transition frequency $f_{t}$, compared to other astrophysical observations.
Specifically, the preference for a higher value of the parameter $f_0$ and the positive-skewed parameters $\alpha_f$ and $\beta_f$ suggest larger differential pair fractions in more massive galaxies, while the preference for a lower value of the parameter $\tau_0$ and the negative-skewed parameters $\alpha_{\tau}$ and $\beta_{\tau}$ indicate shorter merger timescales in more massive galaxies, and the preference for a higher value of $M_{*}$ corresponds
to a higher normalization between the galaxy bulge mass $M_{\text{bulge}}$ and SMBH mass $M$.
The above parameters entirely contribute to the observed relatively high amplitude of the SGWB spectrum ($\Omega_{\rm{GW}}=0.93^{+1.17}_{-0.41}\times 10^{-8}$ at $\text{yr}^{-1}=1/\text{year}$). Note that the posterior distributions of these  parameters are very similar to those reported in \citep{Middleton:2020asl}, where the NANOGrav 12.5-year data set was used. This is because the spectrum amplitudes in both data sets are statistically consistent. On the other hand, the parameters $e_0$ and $\zeta_0$, which determine the shape of the SGWB spectrum, display sharp contrasts in the posteriors obtained from the two data sets, mainly due to the fact that the two data sets are not fully statistically consistent in their measurement of spectrum shape. For the NANOGrav 15-year data set, the distribution of $e_0$ indicates that SMBHs exhibit a large initial eccentricity when transitioning into the GW-emission dominated process, while the larger value of the parameter $\zeta_0$ implies that massive galaxies have, on average, higher densities than what is suggested by a standard Dehnen profile \citep{1993MNRAS.265..250D}.

\textbf{Implication for space-borne GW detectors.}
In the PTA frequency band, SMBHBs are in the inspiral phase, and their radiation power can be calculated using \Eq{dedf1}. After a prolonged period of mutual inspiral, these binaries gradually transition to more circular orbits and enter the merge and ringdown phase, characterized by GW radiation described by \Eq{dedf2} and \Eq{dedf3}. Some of these black holes undergo merger and final coalescence at higher frequencies, entering the space-borne GW detector frequency band. Now we can deduce the properties of the SMBHB population from the PTA results, and further combine \Eq{ogw2} with Eqs.~(\ref{dedf1}-\ref{dedf3}) to obtain the full SGWB spectrum generated by SMBHBs spanning both the PTA and LISA/Taiji/TianQin frequency bands, as depicted in \Fig{spect}.

\begin{figure}[htbp]

\centering
\includegraphics[width=3in]{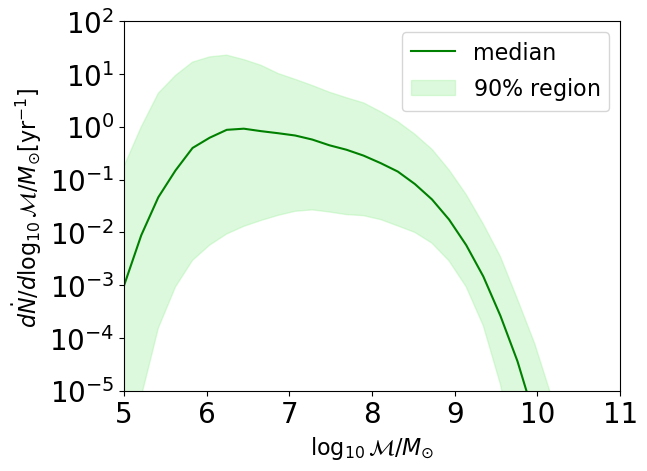}
\caption{
Posterior distributions of the SMBHB merger rate per year per unit logarithmic chirp mass over the range $\mathcal{M} \in [10^5, 10^{11}]\Msun$ where the redshift is integrated over $[0, 5]$. The solid line is the median and the shaded region represents the $90\%$ credible region.}
\label{fig:distribution}
\end{figure}

The relation between merger rate and SMBHB chirp mass is illustrated in \Fig{fig:distribution}. Noted that the merger rate can be obtained from Eq.(2) in \citep{Steinle:2023SMBHBLISA} based on our population model. Essentially, the GW frequency of merging SMBHBs with total masses of $10^4 - 10^7 \Msun$ falls squarely within space-borne GW detector’s bandwidth in the late inspiral, merger and ringdown phase of the binary evolution \citep{Colpi:2019observation}. The total mass ranges approximately correspond with $10^{5} - 10^{7} \Msun$ in chirp mass. Based on the posterior we obtained, the  merger rate of SMBHBs with chirp masses in the range $10^{5} - 10^{7} \Msun$ is estimated as  $\mathcal{R} \simeq 0.54_{-0.52}^{+13.05}$~yr$^{-1}$.

We need to emphasize that the general consensus for the GW detection of the cosmic history of SMBHBs is that PTAs primarily detect the SGWB from the ensemble of the SMBHB population, while LISA/Taiji/TianQin is expected to detect the final coalescence stage of individual systems. However, during the initial stages of the detector in operation, we cannot directly resolve individual sources, and it is reasonable to consider these sources as constituting an SGWB. In fact, as depicted in \Fig{spect}, the spectrum of the SGWB is sufficiently strong that it is very likely to be detected very soon once the detectors are in operation. 


\textbf{Summary.} 
In this letter, we utilize the NANOGrav 15-year data set to constrain the parameters in the astro-informed formation model for SMBHBs, implying that the SMBHBs tend to have a large initial eccentricity in the transition phase between interaction domination and GW domination. The SGWB spectrum from SMBHBs is extrapolated from the PTA frequency band to the space-borne GW detector frequency band. Our results indicate that the forthcoming LISA/Taiji/TianQin will likely detect such an SGWB from SMBHBs in the near future.

By incorporating the evolution of SMBHBs in galaxies with realistic property distributions, we combine astrophysical observations from luminosity and SGWB from PTAs. The GSMF function we used remains valid until $z < 8$ and forms a solid foundation for our entire model. The posterior on GSMF aligns with the prior obtained from \citep{Conselice:2016GSMF,Chen:2018galaxy18}. The constraints on the galaxy pair function and merger timescale are similar to those in \citep{Middleton:2020asl,Steinle:2023SMBHBLISA}. 
Noted that the merger timescale characterizes the galaxy merger originally. Here, we assume that any additional delay between the galaxy merger and SMBHBs merger will be absorbed in the parameter described by \Eq{timescale} as stated in \citep{Chen:2018galaxy18}.


Note that the astro-informed formation model we used is somewhat idealized. Our result is inferred by a particular black hole-galaxy population model which predicts a high IMBH number density. 
To obtain an accurate SGWB energy spectrum, we need a more sophisticated population model. 
Although there is still much work to be done, our work represents an important step forward in this endeavor.


\textit{Acknowledgements.}
We acknowledge the use of HPC Cluster of ITP-CAS. QGH is supported by the grants from NSFC (Grant No.~12250010, 11975019, 11991052, 12047503), Key Research Program of Frontier Sciences, CAS, Grant No.~ZDBS-LY-7009, CAS Project for Young Scientists in Basic Research YSBR-006, the Key Research Program of the Chinese Academy of Sciences (Grant No.~XDPB15). 
ZCC is supported by the National Natural Science Foundation of China (Grant No.~12247176 and No.~12247112) and the China Postdoctoral Science Foundation Fellowship No.~2022M710429.


\bibliography{refs}

\end{document}